\documentclass[twocolumn]{aa} 
\usepackage{graphicx}
\usepackage{amsmath}
\usepackage{txfonts}
\usepackage{array}
\usepackage{hyperref}  
\usepackage{lineno}
\usepackage{CJK}

\def\rhocnm{\rho_{\rm CNM}}
\def\mhcm{\rm m_{H}~cm^{-3}}
\def\msun{M_{\odot}}
\def\msunyr{{\rm M_{\odot} yr^{-1}}}

\def\vw{v_{\rm out}}  
\def\epe{\epsilon_{\rm e}}
\def\epb{\epsilon_{\rm B}}

\def\Mw{\dot{M}_{\rm out}}
 
\def\ergs{{\rm erg~s^{-1}}}

\def\be{\begin{equation}}
\def\ee{\end{equation}}

\begin{document}
\title{Numerical Studies on the Radio Afterglows in TDE: Bow Shock}
\authorrunning{Mou \& Shu}
\titlerunning{Radio Afterglows -- Bow Shock}

\author{ 
Guobin Mou  (\begin{CJK}{UTF8}{gbsn}牟国斌\end{CJK}) \inst{1} 
      \and
 Xinwen Shu  (\begin{CJK}{UTF8}{gbsn}舒新文\end{CJK}) \inst{2}
    }
       
\institute{Department of Physics and Institute of Theoretical Physics, Nanjing Normal University, Nanjing 210023, China; 
     \email {gbmou@njnu.edu.cn}
     \and  Department of Physics, Anhui Normal University, Wuhu, Anhui, 241002, China \
  }

\date{ }

\abstract
{
The origin of radio afterglows or delayed radio flares in tidal disruption events (TDEs) is not fully understood. They could be generated either by a forward shock propagating into diffuse circumnuclear medium (CNM),  or a bow shock around a dense cloud, each of which is fundamentally different.  
To elucidate the distinctions between these two scenarios, we conducted two-fluid simulations incorporating relativistic electrons to investigate the spatial evolution of these electrons after being accelerated by shock. Based on their spatial distribution, we performed radiative transfer calculations to obtain the synchrotron spectra. In Paper I \citep{mou2025}, we reported the results for the forward shock scenario; in this article, we focus on the bow shock scenario.  
{Compared to that from the forward shock whose peak frequency typically lies around GHz and decreases with time, the radio emission from the bow shock peaks at higher frequencies, typically $\sim$1-20 GHz, and its flux rises more steeply than $t^4$ across our explored parameter space. } 
The radio flux from the bow shock also responds to fluctuations in the outflow. The combined effects of the bow shock and forward shock substantially alter radio spectra, causing significant deviations from the single-zone emission model, and in some cases producing {multi-component feature} in spectra. This study highlights the importance of the bow shock, and inspires a novel approach for probing dense gas on sub-parsec scales in galactic nuclei by decomposing the bow shock radio spectrum to reveal the conditions of circumnuclear dense gas. 
}  
 
 
\maketitle

\section{Introduction}
{When a star passes within the tidal radius of a massive black  hole, the star is torn apart in a tidal disruption event (TDE, \citealt{hills1975, rees1988}). }
TDEs have been proven to be capable of producing powerful and fast outflows 
{during the violent self-interaction induced general relativistic apsidal precession \citep{sadowski2016, lu2020}, or during the subsequent super-Eddington accretion phase \citep{dai2018, curd2019}. These outflows could eject a substantial fraction of the disrupted stellar mass, with kinetic energies up to $10^{52}$ erg \citep{lu2020}, and kinetic powers of $10^{44-46}$ erg s$^{-1}$ \citep{dai2018, curd2019}.
Observationally, the presence of ultrafast outflows in TDEs is confirmed directly in UV and X-ray band \citep{blanchard2017, kara2018, blagorodnova2019, hung2019}, 
and the high kinetic energy of winds (large-opening-angle outflows) or jets (collimated outflows) has been directly or indirectly inferred by radio emissions for some TDEs (e.g., \citealt{mattila2018, alexander2020}). } When the outflow interacts with the hot diffuse CNM, a forward shock is generated, while the outflow collisions with dense clouds result in the formation of bow shocks. 
The distinction between these two types of shocks is far more fundamental than simply the difference in CNM or cloud density. 
A forward shock propagates through the CNM with a continuously increasing radius and is sensitive to the density of the CNM. In contrast, a bow shock forms as a nearly stationary shock front enveloping the windward side of a cloud, sweeping through the outflow itself and being primarily sensitive to the ram pressure of the outflow and the covering factor of clouds \citep{mou2021b, mou2022}. 

Shocks are considered to be efficient accelerators of charged particles
{via the first-order Fermi acceleration \citep{drury1983}}, and therefore naturally expected to produce radio emission via synchrotron radiation
\citep{chevalier1998, granot2002, duran2013}. 
{There are some particle-in-cell simulations on the shock acceleration of electrons in the context of supernova remnants (e.g., \citealt{park2015, xu2020}), suggesting that the shock acceleration efficiency of electrons depends on the angle between the background magnetic field and shock normal, and Mach numbers. 
However, these simulations adopt a reduced proton-to-electron mass ratio and are restricted in one dimension, which introduces uncertainties when extrapolating quantitative results to realistic conditions in TDEs. These simulations also do not directly constrain the magnetic field energy $E_B$ in the post-shock gas. 
For these reasons,} the ratio between the magnetic field energy $E_B$ and the energy of relativistic electrons $E_e$ remains largely uncertain. 
The equipartition energy method (\citealt{duran2013}), however, indicates that, within a broad range of ratios between $E_B$ and $E_e$, both the total nonthermal energy ($E_B+E_e$) and the size of the emission region exhibit little variation. 

A gratifying fact is that, during the past decade, an increasing number of TDEs have been found to exhibit radio afterglows, which typically appear months to years after the optical outbursts (e.g., \citealt{alexander2020}). 
Moreover, the recent statistical results indicate that radio afterglows are commonly present in TDEs \citep{cendes2024}. 
{The physical origin of these radio afterglows remains hotly debated. Several competing scenarios have been proposed, including delayed launching of TDE outflow \citep{horesh2021, cendes2022, alexander2026}, off-axis relativistic jets whose beamed emission becomes visible as the jet decelerates \citep{giannios2011, lei2016, matsumoto2023}, outflow encountering transited radial density profile of CNM \citep{matsumoto2024}, bow shocks generated in outflow -- cloud interactions \citep{mou2021b, mou2022, bu2023}. Apart from the bow shock model involving outflow-cloud interactions, 
all other models involve forward shocks in which the shock front sweeps up the surrounding gas. Consequently, the standard \citep{duran2013} or generalized \citep{matsumoto2023} equipartition method can be applied to estimate the fundamental physical parameters of the shocks, including their radii and non-thermal energies.  } 
Many observational studies adopt this {equipartition energy} approach to infer the outflow's kinetic energy and the density--distance profile of the CNM, in which the shocks are equated with forward shock by default \citep{alexander2020, goodwin2022, cendes2024}. 

However, a crucial prerequisite -- determining whether the shock is indeed a forward shock -- should not be overlooked. 
We have noticed some TDEs whose inferred shock parameters potentially exhibit inconsistency with the forward shock scenario. 
For instance, phenomena such as recurrent fluctuations in the radio flux (e.g., AT2020vwl, \citealt{goodwin2024}; 
{see \citealt{zhuang2025} for more instances}), and a decreasing shock radius at some epochs (e.g., AT2018cqh, \citealt{yang2025}) are challenging to reconcile with the forward shock model. 

Given the fundamentally distinct nature of bow shock and forward shock, their predicted radio emissions are expected to exhibit significant differences. 
As shown in \citet{mou2022}, a key feature is that the radio flux is proportional to the covering factor of the clouds being impacted by the outflow. 
This implies that the radio flux can rise sharply when the outflow begins to interact with a cloud, and may fluctuate as the outflow encounters multiple clouds.
Within this framework, the shock parameters inferred using the equipartition method essentially reflect those of the bow shock, rather than the forward shock. 

Current theoretical studies of the bow shock in TDEs mainly rely on analytical models \citep{mou2021b, mou2022, zhuang2025}, which often simplify the problem and overlook complex hydrodynamic processes. 
{The post-shock flow in a bow shock is intrinsically non-uniform. Owing to the curved shock geometry and the subsequent expansion of the shocked outflow around the cloud, both the ram pressure and the magnetic field strength vary significantly throughout the downstream. Consequently, the downstream magnetic field and CRe distribution are highly inhomogeneous. Moreover, the bow shock is not spherically symmetric, so the synchrotron emission is inherently anisotropic and depends on the viewing angle. These complexities make an analytical treatment of the synchrotron emission challenging and motivate our numerical approach. }
Thus, hydrodynamic simulations incorporating cosmic ray electron (CRe) components are essential for understanding the radio spectra in this scenario. 
{This work is a companion to \citet{mou2025}, in which we reported the corresponding simulation results for the forward shock. Here we mainly focus on the bow shock. We find that the two kinds of shocks produce distinctly different radio signatures, including both peak frequencies and light curve behaviors, and we also present the spectral shape when both shocks coexist. }

We introduce the physics related to the models and the settings of the simulations in Section 2 and 3, respectively. Results and discussions are shown in Section 4 and we give a brief summary in Section 5.

\section{Physics of the Models} 

\subsection{Circumnuclear Medium (CNM)}
CNM may {contain} not only hot component 
{with a temperature of $\gtrsim 10^7$ K}, but also dense component manifested as clouds
{(such as clumps making up the torus in active galactic nuclei)}, disks \citep{miyoshi1995}, or spirals {(such as mini-spiral in the Galactic center, \citealt{zhao2009})}. 
For quiescent {supermassive black holes} (SMBHs), the existence of circumnuclear dense structures located within sub-parsec scales from the black hole remains largely unknown. 
One method to confirm their existence is through infrared echoes \citep{lu2016}, which depends on the dust content and only a few cases have been identified so far. Meanwhile, such structures have been observed in the Galactic center -- the mini-spiral (\citealt{tsuboi2016, wang2020}). 
This suggests that circumnuclear dense structures could exist around more quiescent SMBHs. 
A bow shock forms when an outflow encounters a dense structure, regardless of whether it is dusty. 

In this study, we explored the interaction between the outflow and a single cloud. The effect of the outflow on multiple clouds can be treated as the sum of the results of individual clouds, provided there is no mutual obscuration among the clouds. More complex cases are beyond the scope of this paper and will be left for future investigation. 

\begin{figure}
\centering
\includegraphics[width=0.99\columnwidth]{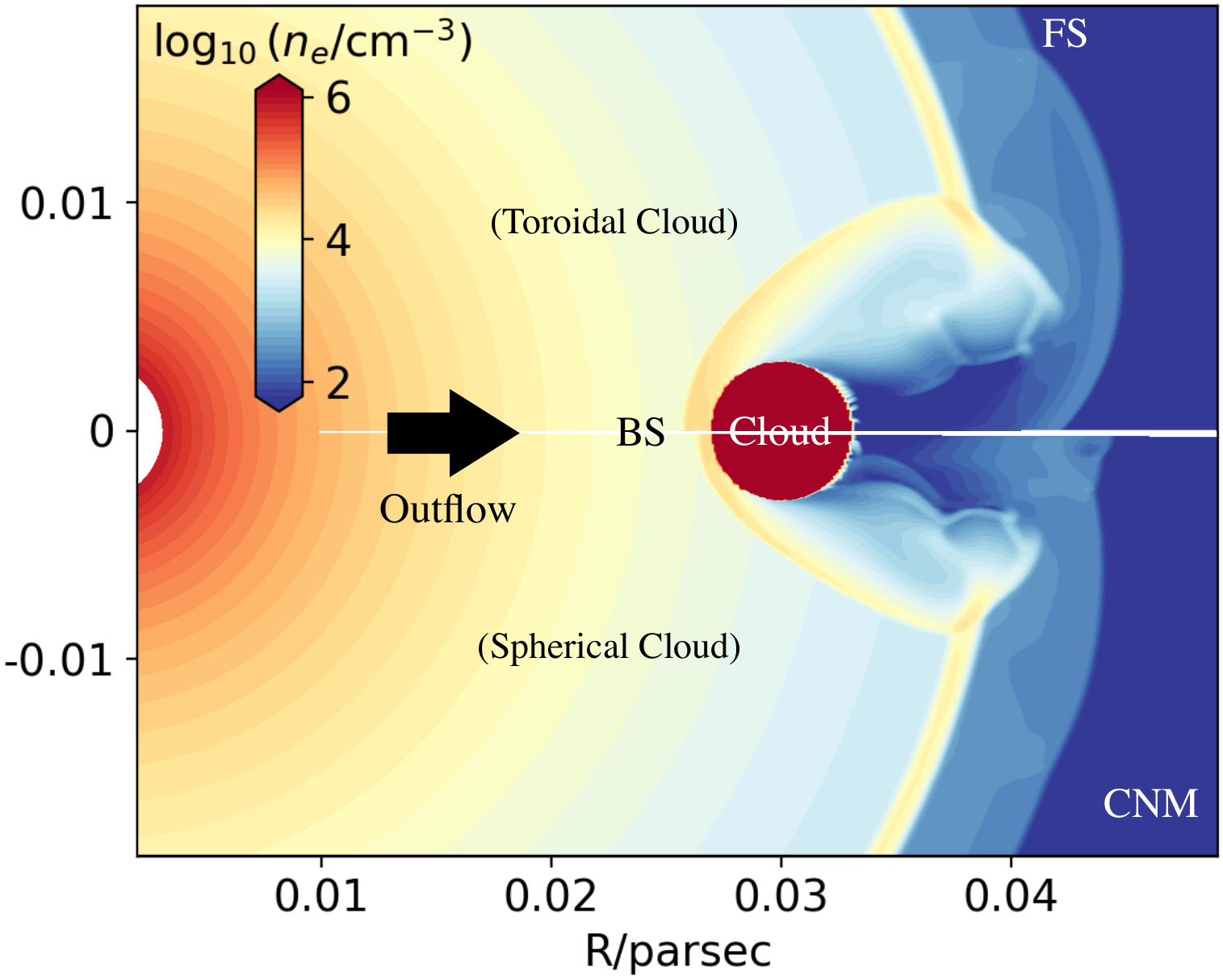}
 \caption{A bow shock formed by outflow impacting a toroidal (model aBt) and a spherical (model bBs) cloud. The bow shock of the toroidal cloud is 2D-shock in nature and appears more extended. {Note that for the spherical cloud model, the cloud is initially placed on the polar axis and deliberately rotated into the equatorial plane to facilitate a direct comparison with the toroidal cloud. } } 
 \label{fig1}
\end{figure}

For the bow shock scenario, the intensity of the radio emission is largely determined by the bow shock energy and the magnetic field strength, which is related to the ram pressure of the outflow. Therefore, the distance between the cloud and the black hole, as well as the size of the cloud, are two key parameters affecting the radio emission. The distance from the cloud center to the SMBH is denoted by $d_c$, and the cloud radius by $R_c$. The specific parameters are listed in Table \ref{tab:tab1}. 
Since the radio flux produced by the bow shock is closely related to the distance of the cloud, we consider two cases for $d_c$: 0.03 pc and 0.2 pc.  
Regarding cloud geometry, both spherical and toroidal clouds are examined, as they influence the bow shock in distinct ways (Figure \ref{fig1}).
For a spherical cloud, the shocked gas expands in three dimensions, resulting in the shock front closer to the cloud surface. In contrast, for a toroidal cloud, the shocked gas expands in two dimensions, forming a more ``extended'' bow shock that is farther from the cloud surface. 
The cloud density is set to $\rho_c = 10^6 ~\mhcm$ for $d_c = 0.2$ pc, and $3\times 10^7~\mhcm$ for $d_c = 0.03$ pc. These values are much higher than the outflow density,  ensuring the cloud remains stable throughout the study. In this paper, the term ``cloud'' refers to a high-density component distinct from the diffuse, keV-temperature CNM. Specifically, ``cloud'' can denote a neutral atomic cloud or partially ionized (\(10^4\)~K) gas. 

The density distribution of the hot CNM is assumed to be similar to that of the Galactic center: $\rhocnm(r)=\rho_0 r^{-n}$, where $\rho_0=3~\mhcm$ and the index $n=1.0$ \citep{gillessen2019}. 

\subsection{TDE outflow} 
TDE outflow could be generated in both self-interaction of bound debris due to GR reprocessing \citep{lu2020}, and the accretion process \citep{dai2018, curd2019}. 
Numerical studies reveal that the outflow could be powerful, and the ratio of mass outflow rate to the mass accretion rate could be considerable (1/10 or more, e.g., \citealt{steinburg2022, thomsen2022}). Accordingly, the mass of the outflow is expected to reach $10^{-2}-10^{-1}\msun$ for disrupting a sun-like star. 

We parameterize the injected outflow with its velocity $\vw(t)$ and mass outflow rate $\Mw(t)$, both of which should be time-dependent.
The outflow is injected from the inner boundary within a period of $\Delta t$ (fixed at 1 yr). 
The parameters of the TDE outflow are quite elusive, and they are closely related to the accretion physics of the TDE which is still poorly understood. 
{Our} setup of the outflow parameters is hypothetical. 
{Here we assumed three simplified cases for outflow: a steady high mass outflow rate, a steady low mass outflow rate, and a highly variable mass outflow rate. In the first two cases, we examined a constant mass outflow rate maintaining for a duration of 1 yr: } $\Mw(t)=\dot{M}_0$, where $\dot{M}_0=0.10~\msunyr$ or $0.02~\msunyr$. The corresponding kinetic energies are $3.4\times 10^{50}$ erg and $1.7\times 10^{51}$ erg, respectively. 
{In the last case, we introduce a random perturbation to the mass outflow rate every 20 days, where outflow density randomly oscillates between 0.02 and 2.0 times the fiducial value of $0.1~\msunyr$. } 
We conducted a preliminary investigation of the spectral characteristics and radio flux variations under this case.

For the velocity, let the maximal outflow velocity be $v_0$ and the minimal velocity be $v_{\rm min}$. The duration of the bow shock (i.e., the outflow-cloud interaction time) is 
\be
t_{\rm BS} \simeq \Delta t + d_{\rm c}/v_{\rm min}-d_{\rm c}/v_0 ~, 
\ee
where $d_{\rm c}$ is the distance of the cloud to the SMBH. This value could be much larger than the outflow's launching time $\Delta t$. 
If the outflow velocity at the late stage $v_{\rm min}$ is low, the lifetime of the bow shock can be significantly prolonged, thereby extending the duration of the bow shock radio emission. The code used is a non-relativistic hydrodynamics code, so we limit the maximum outflow velocity to 0.2c, a value that has been observed in TDEs \citep{kara2018}. In this case, the difference between the Newtonian and relativistic kinetic energies is within 3\%, which does not affect conclusions. 
We assume that the outflow velocity at the inner boundary linearly decreases with time: 
\be
\vw=v_0 (1-\alpha t/\Delta t) 
\ee 
where $t\leq \Delta t\equiv1$ yr, $v_0=0.2c$, $\alpha=2/3$ is a constant for adjusting the velocity of the outflow tail.  

The morphology and properties of the bow shock are closely related to the parameters of the outflow. Due to the high uncertainty and diversity of TDE outflow parameters, we remind that the results of the bow shock scenario are preliminary.

\subsection{Relativistic Electrons and Magnetic Field}

The head of the bow shock can be roughly regarded as a normal shock, while the remaining part of the bow shock corresponds to an oblique shock. When the Mach number is much greater than 1, the post-shock pressure is $P_2\simeq \frac{3}{4}\rho_1 v^2_1 \sin^2 \psi$, where $\rho_1$ is the density of the pre-shock gas and $\psi$ is the angle between the upstream velocity and the shock front, \citealt{landau1987}). Obviously, the bow shock can efficiently dissipate the kinetic energy of the incoming flow. 
Electrons/protons are efficiently accelerated at the bow shock front, and the magnetic field pressure could be amplified to several 
{percent} of the ram pressure due to resonant streaming instability excited by relativistic particles \citep{volk2005}.  

The electron acceleration efficiency $\epe$ is usually assumed to be the fraction of the energy flux that can be dissipated at the shock surface (i.e., the change in the kinetic energy flux across the shock) channeled into the relativistic electrons in the downstream. 
In the frame of shock front, the expression of $\epe$ is 
\be
\epe=\frac{e_2 v_{d}}{\frac{1}{2}\rho_{i} v^3_s (1-C^{-2})}
\ee
where $e_2$ is the energy density of CRe in the downstream, $v_{d}$ is the downstream velocity, $\rho_{i}$ is the pre-shock density, and $C=4$ is the compression ratio. The expression of $\epe$ can be further simplified to $0.6 e_2/e_{d}$, where $e_{d}$ is thermal pressure in the downstream. 
Here we adopt $\epe=0.03$. 
The magnetic field is not included in the simulations; instead, we assume that the ratio of magnetic field energy density to that of CRe in each mesh cell remains constant, i.e., $(8\pi)^{-1}B^2/e_2=\epb/\epe$, where $\epb$ is set to be 0.10. 

The treatment of relativistic electrons and magnetic field strength in this work is exactly the same as in Paper I. For more details, please refer to Sections 2.3 and 2.4 of Paper I.

\section{Methods}

\subsection{Numerical settings}
We conducted the two-fluid simulation with ZEUS-3D code (\citealt{clarke2010}). 
{Here, the term ``two-fluid" refers to thermal gas + CRe, where the CRe component is treated as the second fluid that evolves via its own energy equation and equation of state, dynamically coupled to the thermal gas. }
The hydrodynamic equations without viscosity are
\begin{gather} 
\frac{\partial \rho}{\partial t} + \nabla \cdot (\rho {\bf v})=0 , \\
\rho \frac{d {\bf v}}{d t} = -\nabla (p_{1}+p_{2}) -\rho \nabla \Phi , \\
 \frac{\partial e_{1}}{\partial t} +\nabla \cdot(e_{1}{\bf v})=-p_{1}\nabla \cdot {\bf v}, \label{hydro3} \\
  \frac{\partial e_{2}}{\partial t}+\nabla \cdot(e_{2}{\bf v}) =-p_{2}\nabla \cdot {\bf v}, 
\end{gather}
where $p_1\equiv (\gamma_1-1)e_1$ is the thermal pressure ($\gamma_1=5/3$), $p_2\equiv (\gamma_2-1)e_2$ is the pressure of CRe ($\gamma_2=4/3$), and $\Phi=-GM_{\rm bh}/r$ is the 
{gravitational} potential where $M_{\rm bh}=5\times 10^6 \msun$.
We did not simulate the acceleration process of electrons, but instead, directly injected the CRe component as the second fluid which is embodied as the energy density $e_2$. 
In ZEUS-3D, the shock front is generally resolved across four mesh zones (see Figure 1 in Paper I). Thus, we inject the CRe at the 4th-mesh -- the last mesh resolving the shock where the parameters of the post-shock gas are stabilized. Subsequently, to ensure energy conservation, we adjust the thermal gas energy density $e_1$ by subtracting the corresponding CRe energy density $e_2$.
In addition, we did not include physical diffusion process for the CRe. 
The magnetic pressure in the downstream is much lower than the gas pressure, 
{if the situation is similar to that in supernova remnants \citep{volk2005}. In such a case,} the rapidly flowing downstream gas tends to drag the magnetic field lines, aligning the field lines roughly with the gas flow. On the other hand, for the CRe responsible for the radio emission concerned here, their energies are below GeV and Larmor radii are less than $10^{-11}$ pc, making their motion frozen onto the magnetic field. Under these conditions, CRe primarily move along with the downstream gas and hardly diffuse into the cloud. 

The simulation domain is 2.5 dimensional spherical coordinates, of which the system is symmetric in $\phi-$direction. 
{The ``2.5D" is 3D in physics, and the hydrodynamic code solves the equations in a three-dimensional framework.}
The computation domain is divided into 2880 pieces in $r-$direction with $dr_{i+1}/dr_{i} = 1.0016$, and 360 pieces in $\theta-$direction. The inner and outer radii in the $r$ direction are set to $3.1 \times 10^{-3} - 0.31$ pc for the close cloud models, or $6.1 \times 10^{-3} - 0.61$ pc for the distant cloud models. To save computational resources, we restrict the $\theta$ direction to the range $0$ to $\pi/2$, which can be reconstructed to the $0$ to $\pi$ domain using mirror symmetry. The high resolution adopted here is sufficient to ensure the results are convergent.

\subsection{Calculating the radio spectra}

{The synchrotron spectra are computed from the simulated CRe energy density and magnetic field strength using the same radiative-transfer solver as in Paper~I. We therefore summarize the procedure briefly and focus on the aspects specific to the bow-shock geometry.

For each output time we reconstruct the three-dimensional emitting structure from the axisymmetric $(r,\theta)$ data by revolution about the polar axis, and resample it onto a Cartesian (XYZ) or cylindrical (R,z) grid aligned with that axis (see Figure 2 in paper I). 
Specifically, for the polar direction, we take advantage of the intrinsic axisymmetry to resample the original 2.5D simulation data onto cylindrical grid, while for the equatorial line of sight, we resample the original data into Cartesian grid. 

Within each cell the CRe distribution and field are taken to be uniform, and the specific intensity is advanced along the line of sight by the formal solution of the transfer equation, 
\be 
I_\nu(i{+}1)=I_\nu(i) ~ e^{-\Delta\tau_\nu(i)}+\frac{j_\nu(i)}{\alpha_\nu(i)}\big[1-e^{-\Delta\tau_\nu(i)}\big] ~,  
\ee 
with $\Delta \tau_\nu(i)=\alpha_\nu(i) \Delta x_i$ being the incremental optical depth of cell $i$ at frequency $\nu$, $j_\nu$ and $\alpha_\nu$ the synchrotron emissivity and self-absorption coefficient. 
We continue the calculation iteratively to obtain the intensity $I_{\nu, N}$ from the final edge cell $x_N$ along the line of sight.
Integrating the emergent intensity $I_{\nu,N}$ over the projected emitting area gives the flux density at luminosity distance $d_L$, 
\be 
F_\nu=\frac{1}{d_L^{2}}\sum_k I_{\nu, N}(k) \Delta A_\perp(k), 
\ee 
where $\Delta A_\perp$ is the projected area of the $k$-th edge cell. 
We evaluate two representative viewing directions: a polar and an equatorial line of sight, for which $\Delta A_\perp$ reduces to $2\pi R \Delta R$ and $\Delta Y \Delta Z$, respectively. 

The synthetic spectra are normalized to $d_L=100$~Mpc ($z=0.023$). }

\begin{table}
\centering
\renewcommand{\arraystretch}{1.15}  
\caption{Parameters for modeling the radio afterglows. Annotations for the model names: the first letters a, b, c... -- sequential identifiers, the second or the third Capitals B -- bow shock, F -- forward shock, t--toroidal cloud, s--spherical cloud. Model cBsv considers a randomly varying outflow density. $C_V$ is the covering factor of the cloud.  } 
\setlength{\tabcolsep}{0.25cm} {
\begin{tabular}{cccccc}
\hline \hline
Model  & ${\rm Shape}$ & $d_c$ & $R_c$ & $M_{\rm out}$     & $C_V$     \\
     &                           &  pc      &   pc      & $M_{\odot}$       &       \\ 
\hline
\multicolumn{6}{c}{\text{Only bow shock:}} \\
\hline
aBt    & Torus   & 0.03 & 0.003  & 0.10 & 0.10        \\ 
bBs   & Sphere & 0.03 & 0.003 & 0.10 & 0.0025     \\ 
cBsv  & Sphere & 0.03 & 0.003 & 0.10 & 0.0025    \\ 
dBt    & Torus   & 0.20 & 0.02  & 0.10 & 0.10       \\ 
\hline
\multicolumn{6}{c}{\text{Only forward shock:}} \\
\hline
eF    & --        & --      & --      &  0.10  & 0      \\ 
fF    & --         & --      & --     &  0.02   & 0      \\  
\hline
\multicolumn{6}{c}{\text{Bow shock + Forward shock:}}    \\
\hline
aBFt   & Torus   & 0.03 & 0.003 & 0.10  & 0.10       \\ 
bBFt  & Torus    & 0.03 & 0.003 & 0.02  & 0.10       \\ 
cBFs  & Sphere  & 0.03 & 0.006 & 0.10  & 0.02     \\ 
dBFs  & Sphere  & 0.03 & 0.006 & 0.02  & 0.02     \\ 
eBFt  & Torus     & 0.2 & 0.02 & 0.10   & 0.10         \\ 
fBFt   & Torus     & 0.2 & 0.02 & 0.02   & 0.10        \\ 
gBFs  & Sphere  & 0.2 & 0.04 & 0.10  & 0.02        \\ 
\hline
\hline
\end{tabular} }  
\label{tab:tab1}  
\end{table}

\section{Results and Discussion}

During the collision process, the bow shock sequentially sweeps through the post-shock CNM, the post-shock outflow, and the unshocked outflow. For clouds located in the inner regions (e.g., $r \sim 10^{-2}$ pc), the column density of the CNM swept by the forward shock is usually much lower than that of the unshocked outflow, and the bow shock primarily interacts with 
{the unshocked outflow} 
However, the situation is different for clouds located in the outer region, where the interaction of the bow shock with the shocked CNM and the shocked outflow begins to play an important role. In addition, factors such as cloud morphology and distribution, as well as the angular distribution and variability of the outflow, all influence the bow shock. Thus, the bow shock is complex and encompasses a wide range of subtypes. It is difficult to comprehensively account for all these factors within a single study. In this work, we present a preliminary investigation, focusing on several key characteristics of the radio emission from the bow shock alone, as well as from the combination of the bow shock+forward shock.  
Note that for the spherical cloud, we considered only one cloud positioned on the $+Z$-axis in the bow shock-only models, while we examine two clouds that are symmetrically positioned on the $\pm Z$/polar-axis in the (bow shock+forward shock) cases. 

\begin{figure*}
\includegraphics[width=0.54\textwidth]{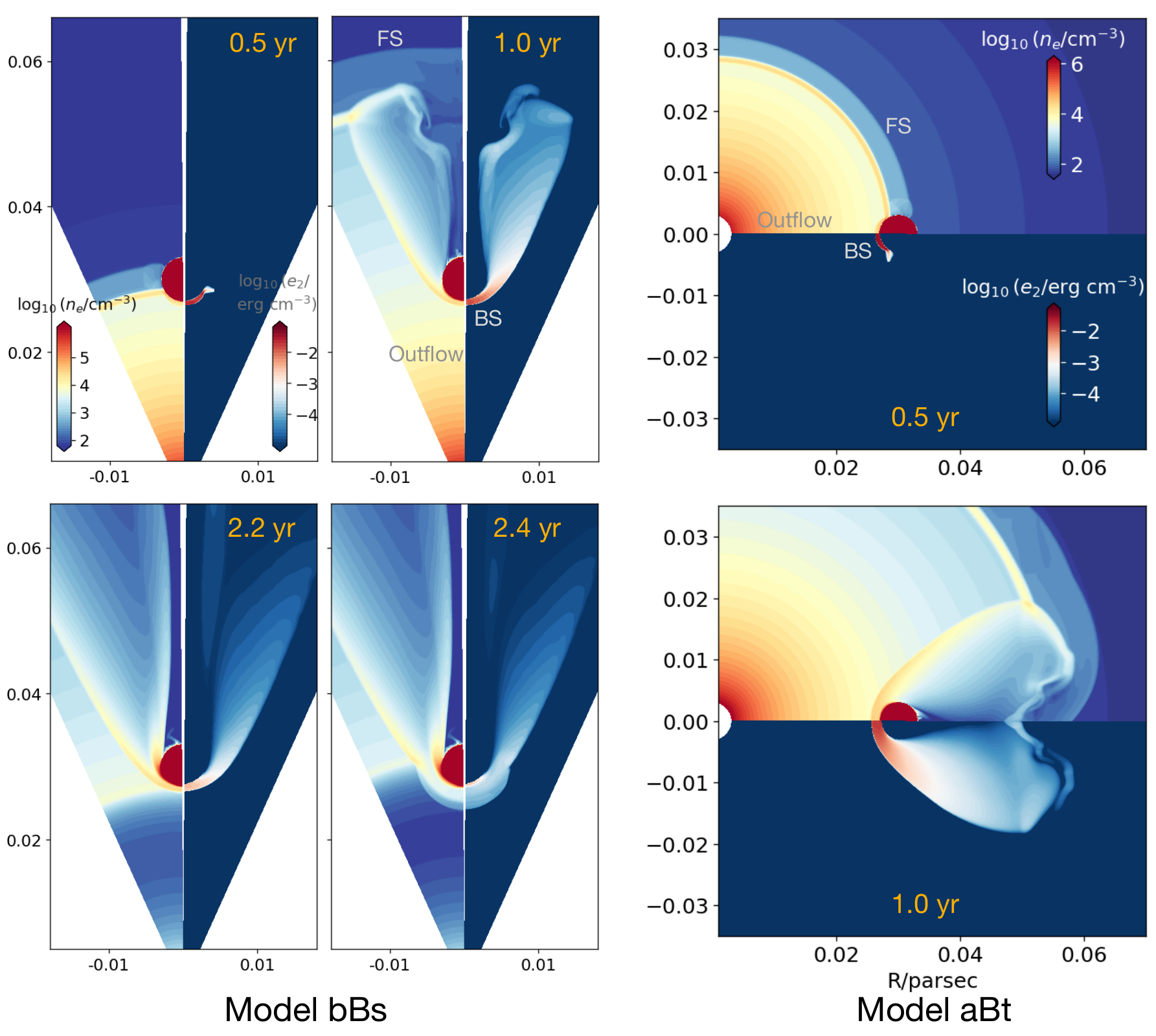}
\hfill %
\includegraphics[width=0.455\textwidth]{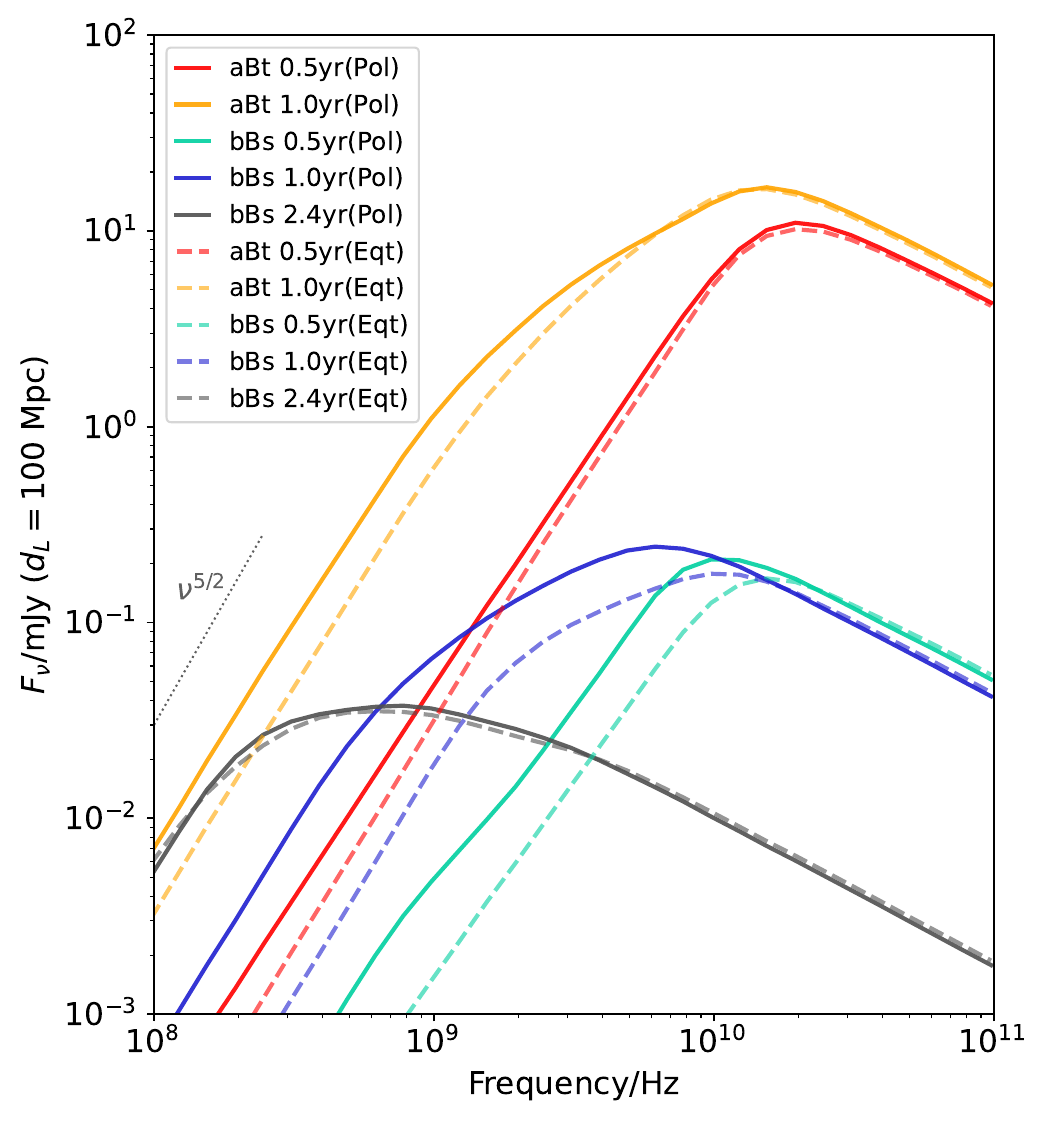}
 \caption{{Left panels: snapshots of pure bow shock model (bBs with a spherical cloud). These four snapshots show the evolution of the bow shock: its onset (as the outflow begins to impact the cloud, $t=0.5$yr), persistent stage, fading stage (as the outflow tail passes through the cloud, $t=2.2$ yr), and disappearance stage (when the outflow has entirely passed the cloud and the outflow-cloud collision ceases, $t=2.4$ yr). 
 Middle panels: snapshots of pure bow shock model (aBt with a toroidal cloud). 
 Right panel: the synthetic radio spectra of bow shocks. The solid and dashed lines represent the polar (Pol) and equatorial (Eqt) lines of sight, respectively.  As the bow shock disappears, both the peak frequency and peak flux significantly decline (model bBs, $t=2.4$ yr). }  }
 \label{fig2}
\end{figure*}

\subsection{Properties of Radio Frequencies for the bow shock} 

{As shown in the left and middle panels in Fig. \ref{fig2}, } the non-thermal components are concentrated at the head of the bow shock, resulting in strong self-absorption of the synchrotron emission. In models aBt and bBs, the peak frequencies are around 10 GHz and change slowly ({right panel in Fig. \ref{fig2}}). As the outflow 
{entirely passes the cloud}, the bow shock also disappears. During this phase, the shocked material rapidly expands, the emission region becomes transparent, and the peak frequency drops to the GHz or sub-GHz range within several months. 

In addition, the bow shock is inherently non-spherically symmetric, and as time goes by, the nonthermal energy in the downstream region of the bow shock develops a jellyfish-like distribution
{(see the snapshots in Fig. \ref{fig2})}. Thus, the spectra from the bow shock are inherently anisotropic. For instance, for the toroidal cloud at the equatorial plane, the self-absorption from the edge-on view is stronger than that from the face-on view, resulting in a higher peak frequency (Fig. \ref{fig2}). In addition, the jellyfish-like distribution of nonthermal component leads to an extra contribution to the low-frequency radio emission, causing the spectra deviating from the conventional form of $\nu^{5/2}$ self-absorption law.

\begin{figure}
\includegraphics[width=0.99\columnwidth]{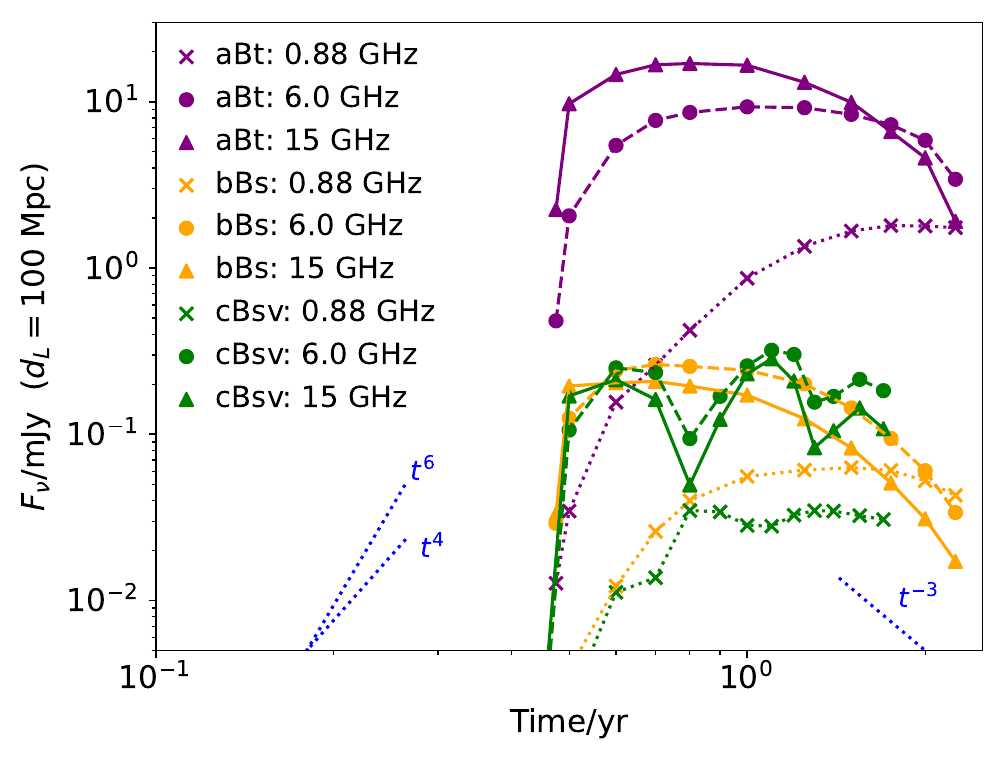}
 \caption{ {Time evolution of} radio flux along the polar/Z direction for pure bow shock models (aBt, bBs, cBsv). {The flux rises sharply as the outflow begins to impact the cloud. Model cBsv adopts a randomly varying outflow density, whereas the other two models employ a slowly varying density while maintaining a constant mass outflow rate and a linearly decreasing velocity (see Section 2.2). As shown, the variability in the outflow density drives corresponding variability in the radio flux. } } 
 \label{fig3}
\end{figure}

\subsection{Basic Levels of the Radio Fluxes} \label{secFlux}

The radio flux produced by the bow shock is influenced by multiple factors, among which the covering factor of the cloud being impacted by the outflow and the ram pressure of the incoming flow at the bow shock play major roles. The covering factor determines the total quantity of CRe, while the ram pressure is closely associated with the strength of the magnetic field.  

The radio flux is positively correlated with the covering factors \citep{mou2022}. 
{This is supported by our simulation results for models aBt and bBs. Specifically,} their covering factors are 10\% and 0.25\%, respectively, differing by a factor of 40, and their radio fluxes differ by a factor of 70 (Figure \ref{fig2}). The toroidal cloud exhibits an exceptionally high flux, owing to the two-dimensional nature and larger size of its bow shock as mentioned above. It is worth noting that the effective covering factor for the bow shock scenario refers to the solid angle subtended by the cloud being impacted, relative to the outflow, rather than to $4\pi$ steradians. Thus, even if the actual solid angle of the cloud is small, the effective covering factor can reach 100\% if the cloud completely blocks a conical outflow or narrow jet. However, such cases are not considered here. 

Our simulations indicate that, when an outflow with a kinetic luminosity of  $10^{43-44}$ erg s$^{-1}$ impacts a cloud located at $10^{-2}$ pc from the SMBH with a covering factor of 10\%, the radio flux can reach 10--20 mJy $(d_L/100 {\rm Mpc})^{-2}$, which corresponds to a peak radio luminosity of $10^{39}~\ergs$. Among the current detected radio afterglows in TDEs, most have peak luminosities below or near this level \citep{cendes2024}. 

\subsection{Variabilities of the Radio Flux}  
Before the outflow reaches the cloud, the bow shock has not yet formed. As the outflow begins to collide with the cloud and the bow shock emerges, relativistic electrons are generated and the magnetic field is amplified, resulting in a sharp rise of the radio emission. 
{In Fig. \ref{fig3}, the flux in early stage increases more steeply than $t^4$, and in some phases even exceeds $t^6$. } Such a delayed and sharp rise in radio flux has been reported in AT2018hyz \citep{cendes2022}, and AT2018cqh \citep{zhang2024,yang2025}. 
Afterwards, with the presence of the bow shock, the radio emission remains relatively stable for a period of $t_{\rm BS}$ 
{(Equation 2). The duration of this stable phase is determined by the outflow--cloud interaction timescale, lasting approximately 1--2 yr for the models shown in Fig. \ref{fig3}. }
Eventually, as the bow shock vanishes, the shocked material rapidly undergoes free expansion, and the flux declines rapidly 
{(steeper than $t^{-3}$ for the models in Fig. \ref{fig3}). Such a rapid decline has been detected in some TDEs. For example, the 5.5 and 9 GHz fluxes of AT2019azh decreased by a factor of 2-3 over 6 months from $\delta t=$ 666 to 849 days \citep{goodwin2022}, consistent with an approximate decline of $\delta t^{-3}\sim \delta t^{-4}$ ($\delta t$ is measured with reference to the estimated outflow launch date). }

Bow shock also responses to the fluctuation of the outflow (e.g., density). In model CBsv, we introduce a random perturbation to the outflow's density every 20 days.  
{This} results in pronounced fluctuations in the radio flux. The {high frequency (6 GHz and 15 GHz)} radio flux exhibits a variation by a factor of {2-4} over a timescale of 2--3 months (Figure \ref{fig3}). 
{This variability arises because the ram pressure of the bow shock responds directly to fluctuations in the outflow density. Since the magnetic pressure is assumed to be a fixed fraction of the ram pressure, the magnetic field strength in the shocked region also varies with the outflow density, leading to pronounced changes in the optically thin, high-frequency radio emission. Consequently, the high frequency radio flux ($\nu \gtrsim \nu_{p}$) is positively correlated with the outflow density. 
In contrast, the low-frequency emission is optically thick and originates primarily from the photosphere where the synchrotron optical depth is approximately unity. 
As the ram pressure decreases, the bow shock expands, causing the photosphere to become slightly larger. Consequently, the low-frequency radio flux tends to increase as the ram pressure decreases, resulting in a weak negative correlation with the outflow density (model cBsv; Fig. \ref{fig3}). 
The pronounced high-frequency} flux fluctuations on a timescale of months have been reported in some sources. {For example, the 11 GHz flux of AT2020vwl exhibited a sharp increase by a factor of 4 over half a year, followed by a rapid decline by a factor of 2 over 3 months (\citealt{goodwin2024}). }

For the forward shock, the nonthermal energy is determined by the accumulated shock energy, rather than by short-timescale fluctuations of the outflow.
{As shown in Paper I, the radio flux above the peak or self-absorption frequency ($\nu>\nu_{p}$) rises relatively slowly as $t^{\Gamma_1}$ where $\Gamma_1 = 3 - 0.25np - 1.25n$ ($n$ is the CNM density index) and decays gradually as $t^{\Gamma_2}$ where $\Gamma_2 \equiv -\frac{3p+3}{2(5-n)}$. For $2 \leq p \leq 3$ and $1 \leq n \leq 2$, the parameters span the ranges of $-1 \le \Gamma_1 \le 1.25$ and $-2 \le \Gamma_2 \le -1$, respectively. } Flux fluctuations may be also produced in certain cases, for example, if there is a break in the index of the CNM density distribution \citep{matsumoto2024}. However, as proven by simulations in Paper I, the flux fluctuations caused by a broken power-law CNM distribution are generally quite weak (on the order of a few percent) and occur on timescales of one year.  

Thus, noticeable amplitude and short-timescale (monthly) fluctuations in the high-frequency flux are another distinctive feature between the bow shock and forward shock scenarios. 
The specific fluctuation amplitude for distinguishing between the two scenarios is difficult to quantify, since it is influenced by several factors including the outflow parameters, contributions from the forward shock or other bow shock, and other effects that can smooth out variations. Based on the study in Paper I, a non-strict criterion of a 20\% amplitude may serve as a useful diagnostic reference. 

{Moreover}, owing to the relatively stable phase in the bow shock radio emission, 
{we would expect that} the presence of multiple isolated clouds would lead to a step-like pattern in the radio flux: a rapid increase over a certain period, followed by a plateau, and then another rise when the next cloud is encountered. 
If the clouds are spatially connected, forming a structure similar to the mini-spirals in the Galactic Center, then as the impact region increases, the radio flux may exhibit a continuous rise. Under such a circumstance, the flux exhibits a sharp initial rise followed by a relatively slower increase. 
The sharp/step-like/{(abrupt early + shallower later)} increase pattern, and rapid decline pattern of the high-frequency flux is a distinctive feature that differentiates the bow shock from the forward shock -- for the latter, we revealed its slowly varying pattern on a timescale of years in Paper I. 

\begin{figure}[htbp]
\centering
\includegraphics[width=0.98\columnwidth]{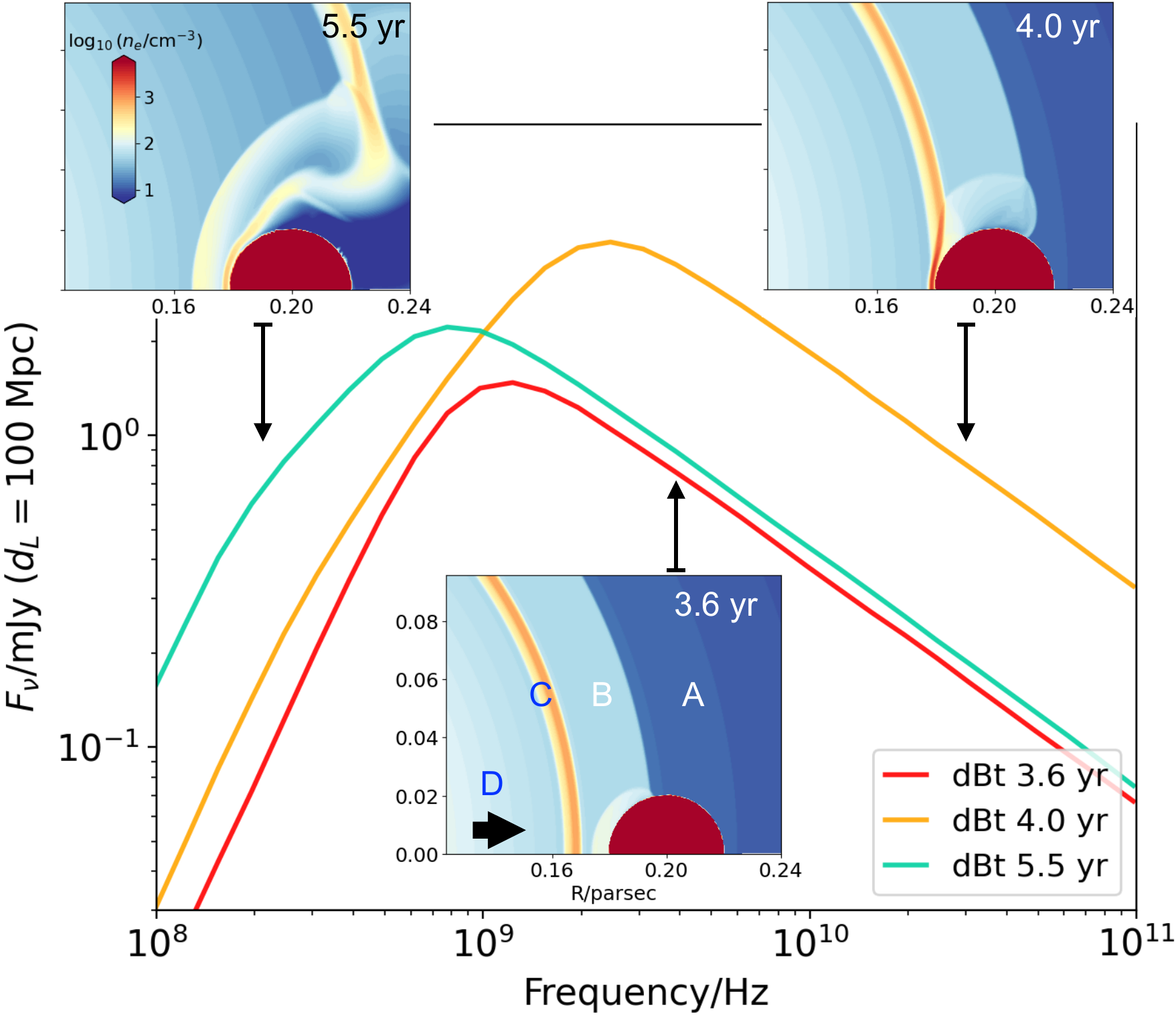}
 \caption{Evolution of the bow shock for a toroidal cloud located far from the SMBH (model dBt with $d_c=0.2$ pc). Because the outflow has been partially processed by the reverse shock before reaching the cloud, the bow shock evolves through three distinct interaction stages. 
The three snapshots, taken at $t=3.6$, 4.0, and 5.5 yr, illustrate the bow shock interacting successively with the shocked CNM, the shocked outflow, and the unshocked outflow. We show the corresponding synthetic radio spectra for these three stages along the polar direction. 
In the snapshot at $t=3.6$ yr, regions A--D denote the CNM, the shocked CNM behind the forward shock, the shocked outflow swept by the reverse shock, and the unshocked outflow, respectively.  }
 \label{fig4}
\end{figure}

\subsection{Influence of the cloud distance $d_c$}
The above summarizes the main characteristics of the bow shocks formed by inner-region clouds. 
When the cloud is located in the outer region (e.g., $r \gtrsim 10^{-1}$ pc), 
{part of the outflow has already been processed by the reverse shock before reaching the cloud. Consequently, the bow shock no longer interacts directly with an unshocked outflow throughout its evolution. Instead, it successively encounters the shocked CNM, the shocked outflow, and finally the unshocked outflow.  

Figure \ref{fig4} illustrates this three-stage evolution using representative snapshots at $t=3.6$, 4.0, and 5.5 yr, together with the corresponding synthetic radio spectra. These snapshots are chosen to highlight the transition of the bow shock through the three interaction stages.  
During the shocked outflow--cloud interaction stage, the relatively denser shocked outflow (with a density up to $\lesssim$4 times that of the unshocked outflow) reaches the cloud, leading to a corresponding increase in the ram pressure of the bow shock. This results in a significantly stronger magnetic field in the downstream of the bow shock, thereby producing a pronounced increase in the high-frequency synchrotron emission. The synthetic radio spectra of $t=3.6$ and 4.0 yr, separated by only 0.4 yr, clearly illustrate this transition. Unlike model cBsv (Fig. \ref{fig3}), where the radio variability is primarily driven by intrinsic density fluctuations in the outflow, this represents an additional mechanism for producing rapid radio brightening. }
 
If the cloud is located even farther away, the outflow may be 
{completely} swept by the reverse shock before reaching the cloud. In this case, the bow shock interacts only with the first two components, which is different from the bow shock in the inner region discussed above. Such cases are not considered in the current study. 

{The radio emission from the bow shocks is sensitive to the cloud distance.} When the cloud is closer to the SMBH, the ram pressure is higher, since the density of the outflow decreases as $d_c^{-2}$. Therefore, for clouds located at different distances, the magnetic pressure in bow shock also decreases approximately as $d_c^{-2}$. For a given high frequency of $\nu_i>\nu_p$, the monochromatic luminosity $L_{\nu_i} \propto B^{(p+1)/2}$ \citep{longair2011}, suggesting that the $L_{\nu_i}$ approximately decreases as $d_c^{-(p+1)/2}$ (assuming an equal total CRe energy). This indicates clouds located in the inner region are more likely to dominate the radio emission, and radio afterglows arisen with relatively shorter time delays are likely to be associated with the bow shock. 
When the cloud is located farther from the SMBH, the overall radio emission is prone to be dominated by the forward shock. However, if the bow shock is present, it could affect the high-frequency spectrum to deviate from a single power-law form.

\subsection{Uncertainty Assessment of the Energy Equipartition Method}
Using the synthetic spectra, we evaluate the reliability of the energy equipartition method \citep{duran2013} for the bow shock scenario.
The geometry of the emission region is unknown to the observer, and it is typically modeled as a thin spherical shell with a thickness of 0.1 times the {shock} radius $r_s$.
{For such a shell-like emitting region, the projected area as seen by the observer is $A=\pi r^2_s$, while the emitting volume is $V=\frac{4}{3}\pi [r^3_s-(0.9r_s)^3]$,} corresponding to a geometry with $f_A = 1$ and $f_V = 0.36$ \citep{cendes2024} \footnote{The definitions of $f_A$ and $f_V$ are the same as in \citep{duran2013}. The observed (projected) area of the emitting shell with a radius of $r$ is $A$, and the total emitting volume is $V$. Then, $f_A \equiv A/(\pi r^2) \leq 1$ and $f_V \equiv V/(\pi r^3) \leq 4/3$. }. With the same parameters $\epe = 0.03$ and $\epb = 0.10$, the shock radius and the total non-thermal energy can be calculated from the synthetic spectrum's $\nu_p$ and $F_{\nu}$ {via the energy equipartition method, denoted as $R_{\rm eq}$ and $E_{\rm eq}$ (\citealt{duran2013}, see also Appendix B in paper I):  
\begin{eqnarray} \label{Req} 
R_{\rm eq} &\approx& \epsilon^{1/17} ~ (1\times10^{17} {\rm cm}) \, [21.8(525)^{p-1}]^{\frac{1}{13+2p}} \,
\gamma_m^{\frac{2-p}{13+2p}} \, \nonumber \\
&&
\times ~ \Big[ F_{\rm p,mJy}^{\frac{6+p}{13+2p}} \, d_{L,28}^{\frac{2(p+6)}{13+2p}}
\, \nu_{p,10}^{-1} \, (1+z)^{-\frac{19+3p}{13+2p}} \Big] \,  \nonumber \\
&&
\times ~ f_A^{-\frac{5+p}{13+2p}} \, f_V^{-\frac{1}{13+2p}}  \,
\Gamma^{\frac{p+8}{13+2p}} \,  4^{\frac{1}{13+2p}} ,  ~~~~(\epsilon \equiv \frac{11\epb}{6\epe})
\end{eqnarray}

\begin{eqnarray} \label{E_e}
E_{\rm eq} &\equiv & E_{e} + E_{B} = \frac{\epsilon_B+ \epsilon_e}{\epsilon_B} E_e~ \nonumber \\
& = &  \frac{\epsilon_B+ \epsilon_e}{\epsilon_B}~ \frac{4 (\gamma_e/\gamma_m)^{2-p} \, 27 c^3 F_{\nu,p}^4 d_L^8 \eta^5 \Gamma^2}{16 \sqrt{3} \pi^3 e^2 m_e^2 \nu_p^7 (1+z)^{11} f_A^3 R^6} \nonumber \\
&\approx&  4  \frac{\epsilon_B+ \epsilon_e}{\epsilon_B} \left(\frac{\gamma_e}{\gamma_m}\right)^{2-p} \times 4.4 \times 10^{50} ~{\rm erg}  \times ~ \nonumber \\ 
& & \quad \Big[ F_{\rm p,mJy}^4 \, d_{L,28}^8 \, \nu_{p,10}^{-7} \, \eta^{5}
\, (1+z)^{-11} \Big] \, \frac{ \Gamma^2}{f_A^{3} \, R_{17}^{6}} .
\end{eqnarray} }
The results are summarized in Table \ref{tab:tab2}. 

Preliminary tests suggest that when this method is applied to the bow shock, the ratio of the estimated emission region radius to the actual value (approximated by the cloud radius) typically falls within the range of 0.3 to 3, indicates an uncertainty of a factor of 3. 
For the nonthermal energy including relativistic electrons and magnetic field, the estimated value is in the range of $0.03\sim 2$ of the actual value. Despite the large uncertainty, we also note that estimates based on early-time radio data (several $R_c/\vw$ after the appearance of the bow shock) tend to be relatively more accurate. This is because the fraction of the non-thermal energy in the jellyfish-like downstream increases over time, while its low energy density results in inefficient radiation. 

{An analytical model of \citet{mou2022} derives the peak frequency $\nu_p$ and peak luminosity $L_{\nu_p}$ of bow shock synchrotron emission directly from the physical parameters of the outflow-cloud interaction: the ram pressure of the outflow at the cloud location which determines the post-shock magnetic field strength and the covering factor of the cloud which determines the total CRe amount. That work is specific to the bow shock picture, while the equipartition method \citep{duran2013} instead infers the emission region size and nonthermal energy from the observed $\nu_p$ and $L_{\nu_p}$ in the context of forward shock picture. The original \citet{mou2022} study} focused on small clouds that are easily disrupted by outflows, whereas the present work examines large and stable clouds. Despite the differences, we note that the analytical results for ${\nu_p}$ and $L_{\nu_p}$ (Equation 12 and 13 {in} that paper) can match the current synthetic spectra (model bBs in Figure \ref{fig2}), provided that the parameter $k_{\rm{bow}}$ in that paper (defined as the ratio of the adiabatic cooling timescale of CRe to $R_c/\vw$) is adjusted from the originally suggested value of 10 to 1. 
{According to the current simulations, we find that the radio emission from the bow shock is dominated by the bow-shock head region (i.e., the bow shock on the outflow-facing side of the cloud). Beyond the head region, the magnetic field in the post-shock region is significantly weaker, so the synchrotron emission from the rest of the bow shock is negligible. Although the adiabatic cooling timescale of the entire CRe population is proven to be as long as $10R_c/\vw$ \citep{mou2021b}, the synchrotron emission is produced primarily by the CRe accumulated at the bow-shock head region, whose the residence or accumulation timescale is $\mathcal{O}(1) R_c/\vw$. Therefore, $k_{\rm{bow}}$ in \citep{mou2022} should be adjusted to 1, which yields a better agreement with the simulation results.} However, for toroidal clouds, the analytical results exhibit significant discrepancies compared to the synthetic spectra from simulations. 

Finally, we did not attempt to estimate the shock velocity, as the approach of dividing $R_{\rm eq}$ by the time delay is entirely inappropriate for the bow shock scenario.

\begin{table}
\centering
\renewcommand{\arraystretch}{1.15}  
\caption{Comparisons of physical parameters in simulations (only bow shock cases) and those inferred from equipartition method.  
{$E_{\rm nth}$ is the real total nonthermal energy from the simulation data, while $E_{\rm eq}$ and $R_{\rm eq}$ are the estimated nonthermal energy and shock radius from the equipartition method. } } 
\setlength{\tabcolsep}{0.09cm} {
\begin{tabular}{cccccccccc}
\hline
\hline
\\[-0.3cm] 
Model   & t & Dirc & $\nu_p$ & $F_{\nu_p}$ & $\lg E_{\rm nth}$ & $R_{\rm eq}$ & $\lg E_{\rm eq}$ & $\frac{E_{\rm eq}}{E_{\rm nth}}$  & $\frac{R_{\rm eq}}{R_c}$   \\
units  & yr     &   --     &  GHz     &mJy  &  & pc & &    \\ 
\hline
aBt  & 0.5  & Pol  & 20  & 11 & 48.0 & 0.0046 & 47.7 & 0.5  & 1.5 \\ 
aBt  & 1.0  & Pol   & 16 & 16  & 48.7 & 0.0068 & 48.0 & 0.2 & 2.3 \\ 
bBs  & 0.5  & Pol & 11 & 0.21 & 45.7 & 0.0013 & 45.9 & 1.6 & 0.4\\ 
bBs  & 0.5  & Eqt & 16 & 0.16 & 45.7 & 0.0008 & 45.6 & 0.8 & 0.3 \\ 
bBs   & 1.0 & Pol & 6.4 & 0.24 & 47.3 & 0.0023 & 46.2 & 0.08 & 0.8  \\ 
bBs   & 1.0 & Eqt & 11 & 0.17  & 47.3 & 0.0012 & 45.8 & 0.03 & 0.4 \\ 
bBs   & 2.4  & Pol & 0.8 & 0.038  & 47.2 & 0.0078 & 46.1 & 0.08  & 2.6 \\ 
bBs   & 2.4  & Eqt & 0.65 & 0.037 & 47.2 &  0.0095 & 46.2 & 0.1 & 3.2 \\ 
dBt   & 3.6  & Pol  & 1.1 & 1.5 & 48.2 & 0.032 & 47.9 & 0.5 & 0.8  \\ 
dBt   & 4.0  & Pol & 2.4 & 4.1 & 48.6  & 0.024 & 48.1 & 0.3 & 0.6 \\ 
dBt   & 5.5  & Pol & 0.8 & 2.2  & 48.8 &  0.053 & 48.2 & 0.25 & 1.3 \\ 
\hline
\end{tabular} }  
\label{tab:tab2} 
\end{table}

\subsection{Combinations of the bow shock and the forward shock}

\begin{figure}
\includegraphics[width=0.49\columnwidth]{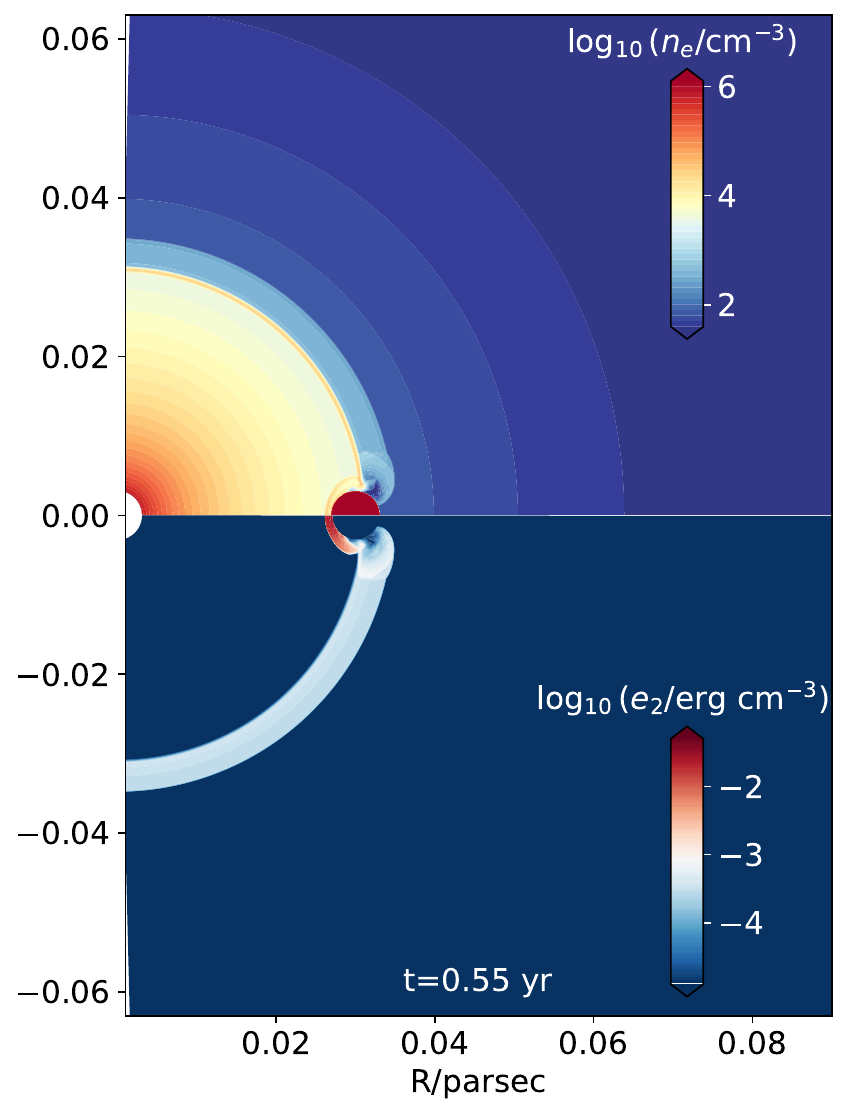}
\includegraphics[width=0.49\columnwidth]{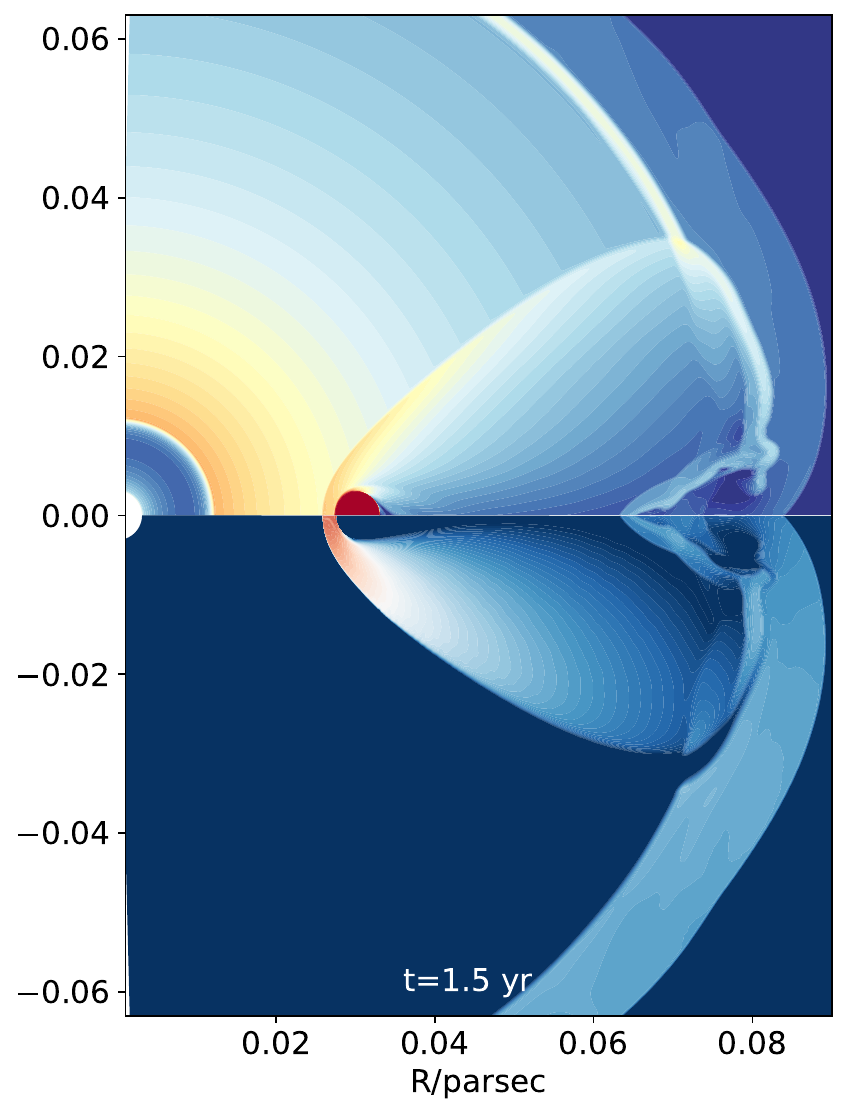}
 \caption{ {Snapshots of model aBFt incorporating CRe accelerated by} both the bow shock and the forward shock ({left: $t=0.55$ yr, right: $t=1.5$yr}). {The upper panels show the density distribution of thermal electrons, and the lower panels show the energy density of CRe.} The left and right panel share the same colorbars.   } 
 \label{fig5}
\end{figure}

\begin{figure}
\includegraphics[width=0.99\columnwidth]{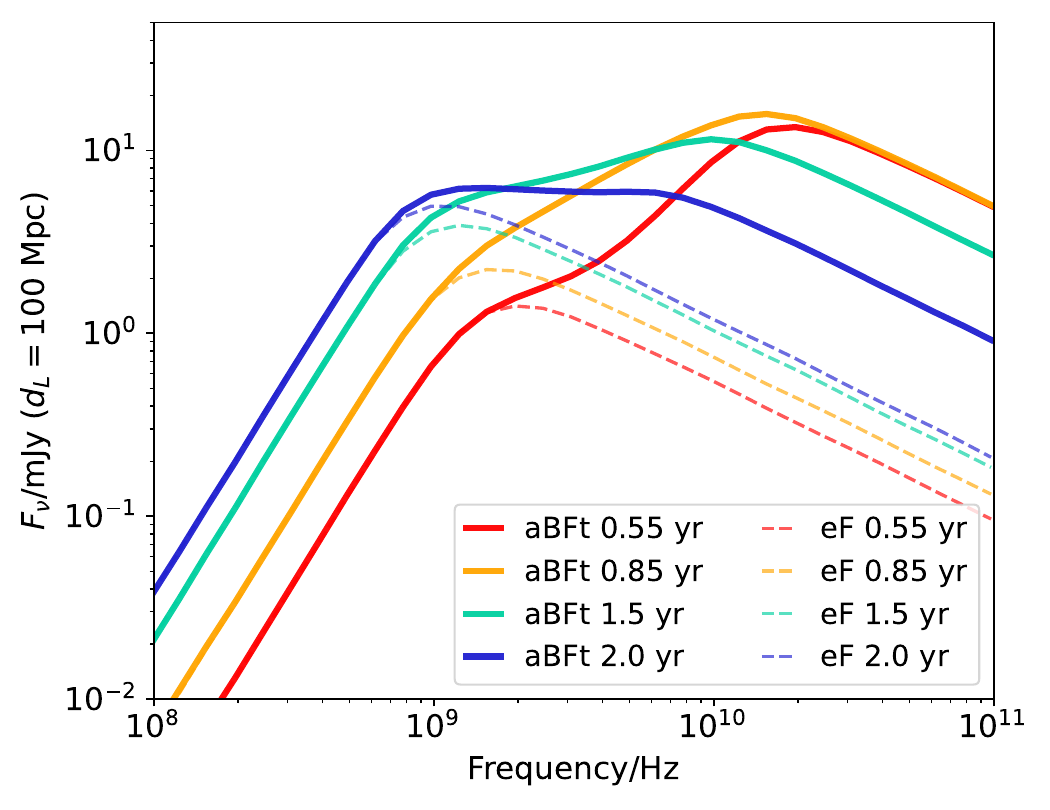}
 \caption{ {Synthetic radio spectra at 4 epochs for the toroidal cloud model in which CRe are accelerated by both the forward shock and the bow shock (model aBFt; solid lines), compared with the corresponding cloud-free, pure forward shock case (model eF; dashed lines). 
The bow shock contribution produces a prominent high-frequency excess above several GHz, manifesting as a double-peaked or multi-component spectral shape that is absent in the pure forward shock case. } }
 \label{fig6}
\end{figure}

{When considering coexisting of bow shock and forward shock, }
the overall radio emission spectrum is determined by the combined contributions of {both shocks} (Figure \ref{fig5}).  
As shown in Paper I, the radio flux from the forward shock increases slowly over time, 
{i.e., $F_{\nu}(t) \propto t^{\Gamma_1}$ for $\nu > \nu_{p}$ where $\Gamma_1 \leq 1.25$ }. 
Consequently, the early-time radio emission from the forward shock is generally weak, 
{driven by this slow temporal growth where $\Gamma < 1.25$ for typical parameters of $p>2$ and $n \geq 1$. }  

For the bow shock, its radio flux is determined by the effective covering factor and the cloud's distance $d_c$ (Section \ref{secFlux}). 
If the cloud is located close to the black hole, say 0.03 pc, its bow shock radio emission may significantly exceed that from the forward shock. 
The bow shock typically dominates the high-frequency emission, while the forward shock dominates the low-frequency emission. 
In model aBFt 
{where both shocks are included}, the high-frequency flux from the bow shock greatly exceeds that from the forward shock by several tens of times at certain times (Figure \ref{fig6}), while the low-frequency of $\nu<\nu_p$ emission is primarily contributed by the forward shock. 
As a result, the spectrum deviates significantly from the one-zone model and may display a double-peaked {or multi-component} feature. 
Furthermore, the low-frequency flux evolves more slowly, while the high-frequency flux varies more rapidly, reflecting the distinct variability characteristics of the two shocks.

\begin{figure}
\includegraphics[width=0.99\columnwidth]{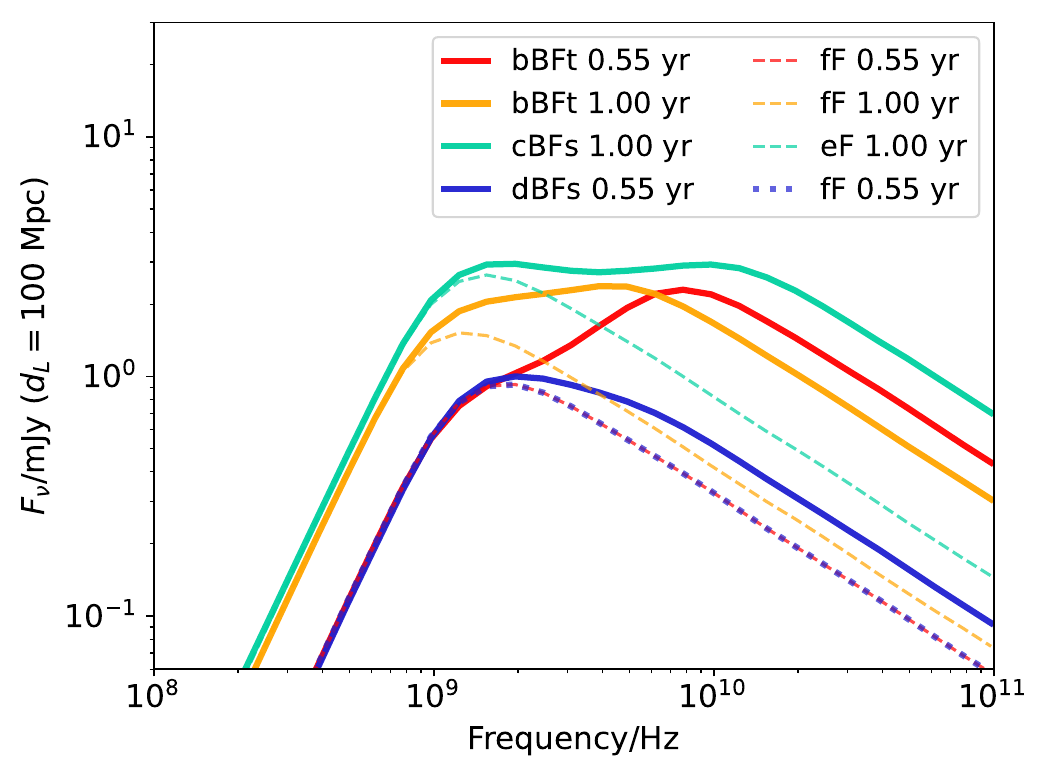}
\includegraphics[width=0.99\columnwidth]{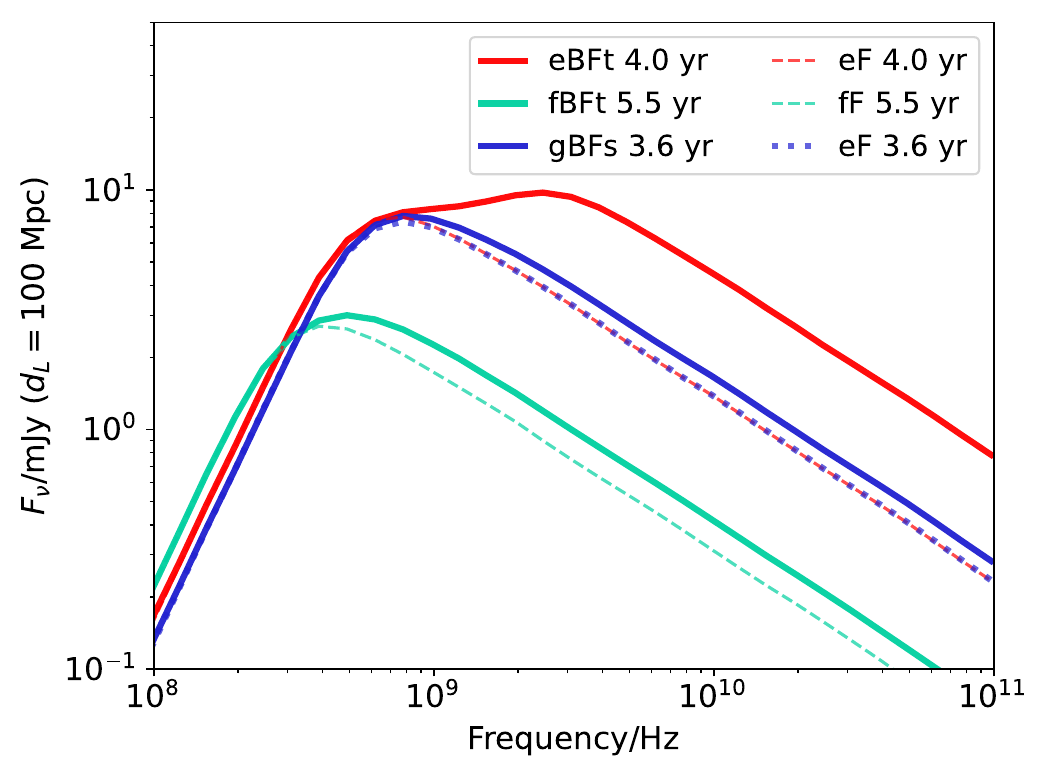}
 \caption{Synthetic radio spectra for 
 {other models in which CRe are accelerated by both shocks. The upper panel shows the close cloud cases with $d_c=0.03$ pc, where the three models differ in cloud shape (covering factor) and mass outflow rate: model bBFt adopts a toroidal cloud with a low mass outflow rate; models cBFs and dBFs adopt spherical clouds, with the former having a high mass outflow rate. The high-frequency radio flux (dominated by the bow shock) exhibits a positive correlation with both the mass outflow rate (comparing bBFt and aBFt) and the cloud covering factor (comparing bBFt and dBFs). Even if the covering factor is small, as long as the mass outflow rate is sufficiently high, considerable bow-shock radio emission can still be produced (model cBFs). The dashed lines represent the comparison spectra without the bow shock. The lower panel shows distant cloud cases with $d_c=0.20$ pc. In such cases, the radio emission from the bow shock becomes appreciable relative to the forward shock only when both the mass outflow rate and the covering factor are high enough. }} 
\label{fig7}
\end{figure}

Figure \ref{fig7} presents the results of other close cloud models ($d_c=0.03$ pc), including cases with lower outflow mass (bBFt) and spherical clouds with low covering factors (cBFs, dBFs). Notably, even in the case where the bow shock expected to be weak (model dBFs, with low-mass outflow and low covering factor), the radio flux from the bow shock can still reach 60\% of the forward shock flux, resulting in a spectral hardening near $\nu_p$. In the other models, the {multi-component feature} are evident.

For the distant clouds, the contribution from the bow shock to the radio emission is relatively weak (Figure \ref{fig7}). For example, in model gBFs (covering factor of 2\%), the bow shock contributes about 20\% of the flux at high frequencies, and the overall spectrum does not differ significantly from the case without the bow shock. In contrast, for toroidal cloud with higher covering factor, the bow shock's contribution could be significant. In model eBFt, the radio flux from the bow shock can be 2-3 times higher than that from the forward shock, resulting in a {multi-component} spectrum. 

These findings suggest that the bow shock could be an important contributor to the radio afterglow. When the bow shock contribution is strong, it may dominate the overall radio emission, or produce a prominent {multi-component} feature, or broaden the synchrotron spectral peak. When its contribution is weak, it may harden the spectral index. Since the radio emission from the bow shock evolves more rapidly than that from the forward shock, these effects are expected to exhibit temporal variability. 

\begin{figure}
\includegraphics[width=0.99\columnwidth]{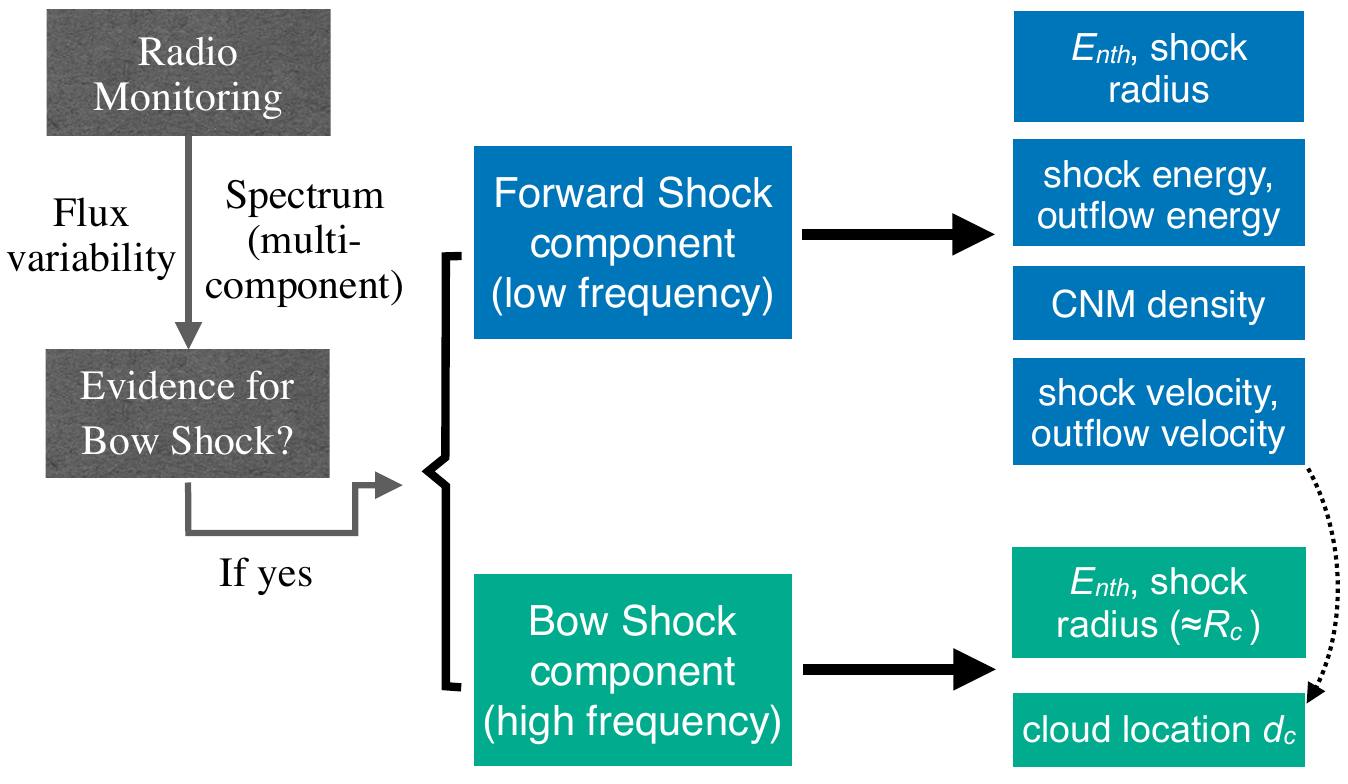}
 \caption{A scheme for exploring the physics of the outflow and CNM environment by multi-frequency monitoring. 
 {Multi-epoch radio spectra allow for the identification of a bow shock based on the rapid flux variability and multi-component spectral features. If confirmed, the spectrum can be decomposed into low- and high-frequency components, representing the forward and bow shocks, respectively. Equipartition analysis of these individual components yields their respective nonthermal energies ($E_{\rm nth}$) and shock radii, which can be further utilized to infer the properties of the outflow, CNM, and clouds. } } 
 \label{fig9}
\end{figure}

The distinct characteristics of the two types of shocks highlight the importance of multi-frequency observations. As mentioned above, the variability characteristics of high-frequency flux (such as rapid rise, fast decline, and fluctuations), as well as the presence of multiple components in the spectrum (manifesting as {multi-component} features), can be used to identify the existence of a bow shock. In Figure \ref{fig9}, we propose a 
{conceptual framework} for exploring the outflow parameters and the physics of the circumnuclear environment when both shocks coexist. 
Specifically, the two- or multi-component radio spectrum (if confirmed) can be analyzed separately, with the high-frequency component corresponding to the bow shock and the low-frequency component to the forward shock. Combining information from both components, it is possible to constrain the outflow parameters and properties of the circumnuclear environment. 
{Applying the equipartition method to the forward shock spectrum yields its shock physics including the nonthermal energy $E_{\rm nth}$ and shock radius, which can be further utilized to infer the shock or outflow velocity, the shock energy and the post-shock CNM mass and density (see Appendix B in Paper I for more details and cautions). 
Similarly, an equipartition analysis of the bow shock spectrum yields its shock size and nonthermal energy, although these parameters can be more precisely determined via two-fluid simulations. This in turn constrains the cloud radius $R_c$($\approx$ bow shock size) and the bow shock energy. Furthermore, the time delay of the bow shock emission, combined with the outflow velocity inferred from the forward shock analysis, can be utilized to estimate the cloud's distance from the SMBH ($d_c$). } 

\begin{figure}
\centering
\includegraphics[width=0.99\columnwidth]{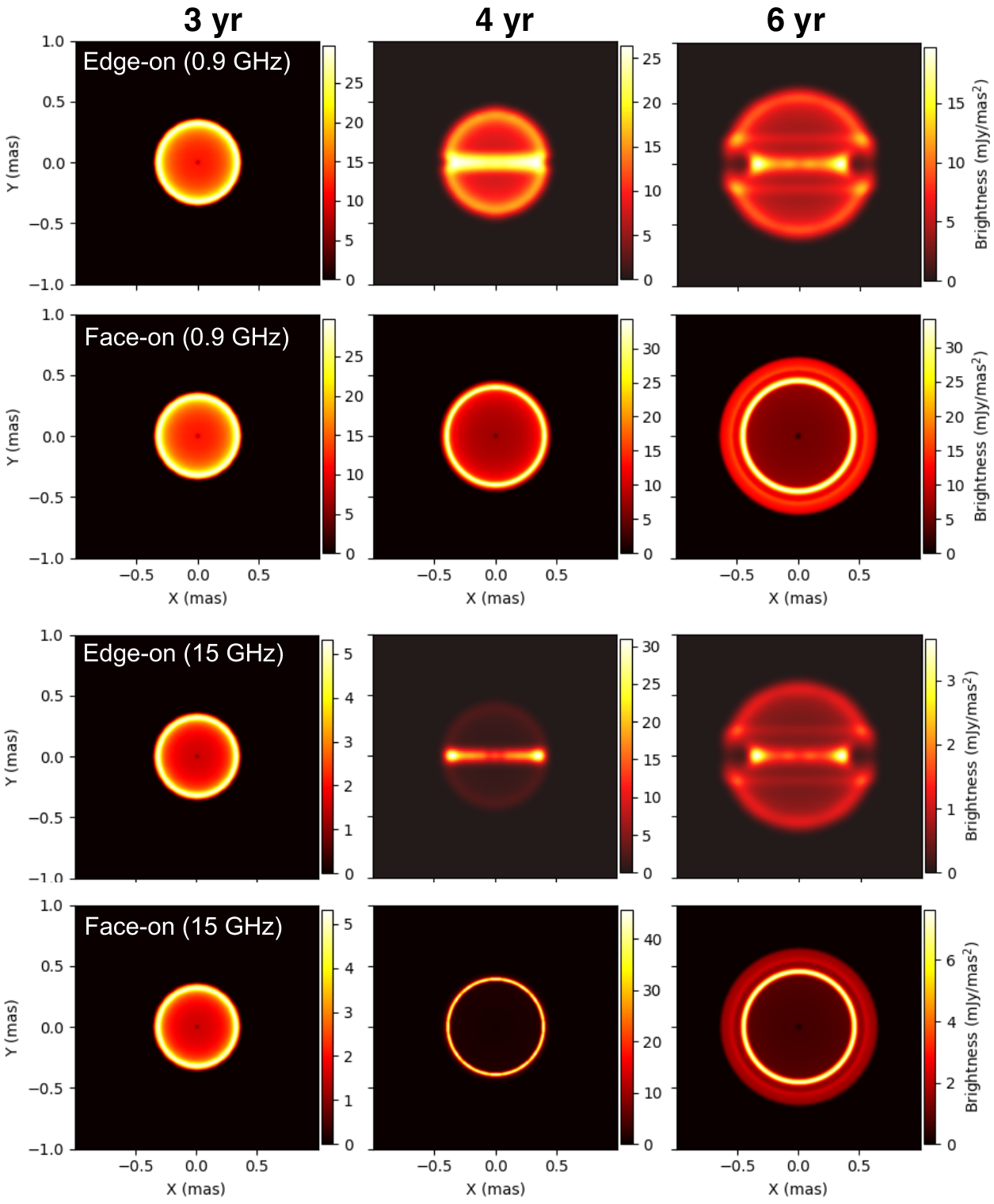}
 \caption{ Synthetic radio maps in $2\times 2$ mas$^2$ for model eBFt (a toroidal cloud located at $d_c=0.2$ pc). The axes are labeled in milliarcseconds ($d_L = 100$ Mpc and $d_A = 96$ Mpc). The three epochs correspond to the phases before the bow shock emerges (3 yr when only the forward shock exists), the early stage (4 yr) and the late stage (6 yr) of the outflow-cloud interaction. } 
 \label{fig10}
\end{figure}

In Figure \ref{fig10}, we present synthetic radio images of model eBFt at different epochs and viewing angles. The radio structure of the forward shock clearly expands continuously over time, while the bright spots (edge-on view) or bright ring (face-on view) from the bow shock remain stationary. 

{At our assumed distance of $d_L=100$ Mpc, the characteristic size of the sub-parsec forward shock and bow shock structures ($\sim$1 mas at 4-6 yr; Fig. \ref{fig10}) falls within the capability of current Very Long Baseline Interferometry (VLBI) networks (e.g., VLBA and EVN), which achieve sub-milliarcsecond angular resolution at GHz frequencies. To date, only a few radio TDEs have been observed with VLBI, and in most cases the radio emitting region remains unresolved, providing only upper limits on the shock size and expansion velocity (e.g., \citealt{yang2016, yang2025}). 
One example is the dust-obscured TDE Arp 299-B AT1  ($d_L = 44.8$ Mpc), for which VLBI monitoring over a decade has resolved changes in the radio emission region on scales of a few mas \citep{mattila2018}. 
This demonstrates that resolving the morphological evolution of radio structures in at least some nearby TDEs is already feasible with current VLBI networks, which could provide a direct diagnostic of the shock scenario. 

In addition, the radio emission from the bow shock is concentrated around the cloud as illustrated in Figure \ref{fig10}. In realistic environments with numerous clouds, bow shocks produce isolated bright knots, arms or rings, reflecting the spatial distribution of the clouds. In contrast, the forward shock generates a coherent bright shell (post-shock CNM) characterized by limb brightening and a relatively faint interior. 
Beyond the evolutionary features discussed above, these morphological characteristics also help distinguish the bow-shock and forward-shock origins of radio emission. Making this distinction also demands high spatial resolution, and is thus suited to nearby TDEs where the clouds are located at parsec-scale distances from the SMBH. }

\section{Conclusions}  

In this study, we explored the bow shock scenario for the radio afterglows in TDEs via two fluid hydrodynamic simulations. 
Simulations provide the spatial distribution of non-thermal components. We reconstruct these data to the three-dimensional space, and calculate the synthetic synchrotron spectra via radiative transfer. This study firstly reveals the radio spectral characteristics of the bow shock and highlights its characteristics and its role in the overall radio emission. The main findings are summarized as follows.  

1. The radio emission from the bow shock arises from quite efficient shock dissipation at the shock front. Compared to the forward shock, the emission region associated with the bow shock is more spatially concentrated with higher nonthermal energy densities and typically exhibits a higher peak frequency {(typically $\sim$1-20 GHz across the explored parameter space) }. 

2. The radio flux from the bow shock increases sharply at the onset of outflow-cloud interaction 
{(steeper than  $t^4$)} and decreases rapidly as the interaction ends {(steeper than  $t^{-3}$)}. This behavior is distinct from the slow evolution pattern of that from the forward shock. 

3. The radio flux of the bow shock is sensitive to the parameters of the outflow; when the outflow undergoes fluctuations, the radio emission from the bow shock also exhibits corresponding variability. This is also different from the forward shock. 

4. For the radio emission from bow shock, estimating the shock radius and nonthermal energy via the conventional energy equipartition method can lead to significant errors. In contrast, estimates based on early-time radio data are relatively more accurate for the bow shock. 

5. When the bow shock and forward shock coexist, the overall radio emission is governed by their relative strengths. 
If the bow shock contribution is significant, it may dominate the overall radio emission, or produce a {multi-component} spectrum. 
If the bow shock is weak, it can still cause spectral hardening at high frequencies. Simulations suggest that for circumnuclear clouds within sub-parsec scales, the bow shock can make a non-negligible and detectable contribution to the overall radio spectrum.

The circumnuclear dense structure near a SMBH as a reservoir is important for triggering the fueling of the SMBH.  
Moreover, these pre-existing structures can significantly affect or collimate the direction of AGN outflows, thereby influencing AGN feedback on galactic scales. 
Our study {presents a conceptual framework for studying} whether circumnuclear dense structures exist around quiescent SMBHs -- utilizing radio emission to {diagnose} the presence and strength of the bow shock. 
Different from the infrared dusty echo \citep{lu2016}, this radio method is not dependent on the abundance of dust, and is therefore more broadly applicable for probing circumnuclear dense gas.  

{This approach is feasible with current radio facilities. Simulations predict a peak bow shock flux of 0.1-10 mJy($d_L$/100 Mpc)$^{-2}$ for representative parameters, placing it well within reach of those radio telescopes used for TDE observations (e.g., VLA, MeerKAT, ATCA) for sources within a few hundred Mpc. Next-generation facilities such as SKA could extend the search distance to several Gpc. 
Moreover, identifying and decomposing the multi-component spectra requires frequency coverage spanning $\sim$0.5--30 GHz. This can be achieved by combining several facilities such as ASKAP (0.7-1.8 GHz), MeerKAT (0.54-3.5 GHz) and VLA (1-50 GHz). 

This approach also has limitations. Identifying and decomposing the bow shock component in radio spectra demands quasi-simultaneous multi-frequency observations, and temporal sampling on a timescale of months to capture the rapid evolution of the bow shock emission. 
In addition, the effects of multiple clouds have not yet been explored and could introduce additional complexity in the spectra. 
We will investigate these issues in future simulations. 

Furthermore, the evolution of the radio-emitting region (a stationary bow shock versus a continuously expanding forward shock) and the morphological characteristics (the bow shock producing isolated bright structures, while the forward shock produces a coherent bright shell with limb brightening) can independently verify the origin of the radio emission. However, this requires VLBI observations with sub-milliarcsecond resolution, which is applicable only to nearby sources ($d_L \lesssim 100$ Mpc) where the clouds are at parsec-scale distances.  }

\section*{Acknowledgements}
We thank the anonymous reviewer for the constructive comments which significantly help improve the manuscript. G.M. was supported by the National Key R\&D Program of China (Grant No. 2023YFA1607904), and the NSFC (nos. 12473013). 
X. Shu is supported by NSFC through grant Nos. 12192220 and 12192221.

\appendix
\setcounter{figure}{0} 
\renewcommand{\thefigure}{A\arabic{figure}}
\setcounter{table}{0}
\renewcommand{\thetable}{A\arabic{table}}

\section{Numerical resolution test}

\begin{figure}
\centering
\includegraphics[width=0.92\columnwidth]{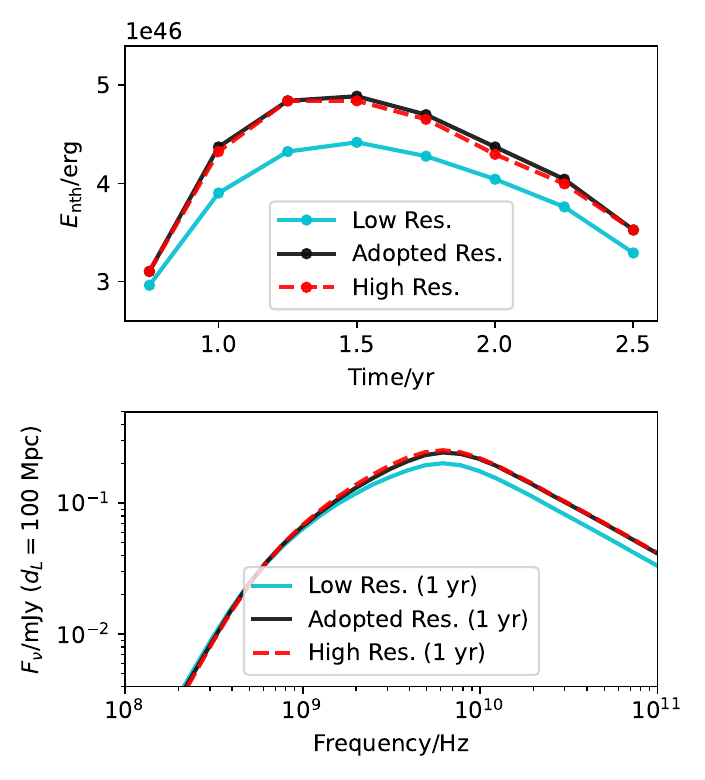}
\caption{Numerical resolution test for model bBs. The top panel shows the time evolution of the total energy of non-thermal electrons accelerated by the bow shock. The bottom panel displays the radio spectra of the bow shock at $t = 1\text{ yr}$, viewed along the polar axis. }
\label{figA1}
\end{figure}

{We have performed a numerical resolution test, and the results demonstrate that the resolution adopted in the main text is sufficient to ensure convergence. In Figure \ref{figA1}, we present the resolution test based on model bBs: the grid cell size of the low-resolution run is twice that of our adopted setup, while the high-resolution run has a cell size half that of the adopted one. Our adopted resolution shows good convergence in both the time evolution of the nonthermal electron total energy and the radio spectra. }

\end{document}